\documentclass[authoryear,5p,times]{elsarticle}
\usepackage{graphicx}
\usepackage{color}
\usepackage{amssymb,amsthm,amsmath}
\usepackage{aas_macros}
\usepackage{bm}

\usepackage{txfonts}
\usepackage[table]{xcolor}
\usepackage{booktabs}
\usepackage{verbatim}
\usepackage{lipsum}
\usepackage{hyperref}
\usepackage{mathtools}
\usepackage{hypernat}
\journal{Astronomy and Computing}
\begin{document}
%
%
\begin{frontmatter}
\title{MEPSA: a flexible peak search algorithm designed for uniformly spaced time series}
%
%
\begin{keyword}
gamma rays: bursts --
methods: statistical --
Design and analysis of algorithms: Pattern matching
\end{keyword}
%

\author[UNIFE]{Cristiano Guidorzi}\ead{guidorzi@fe.infn.it}

\address[UNIFE]{University of Ferrara, Department of Physics and Earth Sciences, via Saragat 1, I--44122 Ferrara, Italy}


\begin{abstract}
We present a novel algorithm aimed at identifying peaks within a uniformly
sampled time series affected by uncorrelated Gaussian noise.
The algorithm, called ``MEPSA'' (multiple excess peak search algorithm),
essentially scans the time series at different timescales
by comparing a given peak candidate with a variable number of adjacent bins.
While this has originally been conceived for the analysis of gamma--ray
burst light (GRB) curves, its usage can be readily extended to other astrophysical
transient phenomena, whose activity is recorded through different surveys.
We tested and validated it through simulated featureless profiles as well as
simulated GRB time profiles.
We showcase the algorithm's potential by comparing with the popular algorithm by
Li and Fenimore, that is frequently adopted in the literature.
Thanks to its high flexibility, the mask of excess patterns used by MEPSA
can be tailored and optimised to the kind of data to be analysed without modifying
the code. The C code is made publicly available.
\end{abstract}

\end{frontmatter}

%
\topmargin -1.3cm

\section{Introduction}
\label{sec:intro} 
Transient phenomena are manifestation of various classes of astrophysical
sources. Their study contributes to characterise the dynamics of such objects
and gain clues on the physical processes responsible for their behaviour
and evolution with time. In the current time domain era,
most if not all transient sources are discovered and signal their transient
character through the emission of one or multiple peaks in their flux time
profile as a manifestation of enhanced activity. In high--energy astrophysics,
for instance, this is the case for black--hole candidates in binary systems
(e.g., \citealt{Remillard06rev}), that of outbursting magnetars
(e.g., \citealt{Mereghetti08rev}), or that of super fast X--ray transients
(e.g., \citealt{Sguera05,Romano14}).

Moving to most energetic and disruptive events on stellar scales, gamma--ray
bursts (GRBs) are no exception. Their gamma--ray time profiles, lasting
from a fraction of a second all the way up to thousands seconds
\citep{Kouveliotou93,Horvath10,Levan14},
are characterised by a variable number of pulses with no firm
evidence for periodicity.
Such pulses often cluster within some emission episodes, separated by
so--called quiescent times, during which the gamma--ray activity drops to
the detector's background level. The so--called GRB ``prompt'',
i.e. the gamma--ray emission itself, is still the least understood aspect
of the GRB phenomenon: e.g., what is (or are) the gamma--ray emission mechanism(s),
at what distance from the collapsing object, which kind of environment
(see \citealt{KumarZhang14rev} for a recent review).
One of the open issues concerns the stochastic (or deterministic?) process
which rules the time profile and its complicated multi--pulse structure
\citep{Greco11}.

To tackle this, one must develop an effective technique to detect as
many peaks as possible in the observed light curves to a reliable degree,
properly accounting for the statistical uncertainties affecting the time series.
Several authors in the GRB literature faced this problem building on
different techniques: \citealt{Li96} (hereafter, LF) set up a simple but effective
algorithm based on the Poisson statistics affecting photon counting detectors.
Other authors made extensive use of LF algorithm \citep{NakarPiran02a,DragoPagliara07,Bhat12}
to study the peak intensity as well as the distribution of waiting times, defined
as the time intervals between adjacent pulses.
In an alternative approach, \citet{Quilligan02} set up a dedicated
filter based on wavelets to suppress the statistical noise in GRB time profiles
and study the basic statistical properties of pulses as a consequence.
More recently, \citet{Charisi14} studied the statistics of waiting times between
emission episodes and precursor activity in GRBs observed by several past and present
experiments by adapting an algorithm originally developed for gravitational data analysis.
This is based on a combined time--frequency decomposition of the time series variance.

Motivated by the peak detection problem in GRB time series, in this paper we propose a new
algorithm, called ``multiple excess peak search algorithm'' (hereafter, MEPSA),
that is described in Sect.~\ref{sec:desc}.
As it will be shown in Sect.~\ref{sec:valid}, compared with the previous LF algorithm
and with a more conservative version of the same devised by us, MEPSA is characterised
by a lower rate of false positive and a higher rate of true positive events, particularly
for low signal--to--noise ratio (SNR) peaks.
Moreover, thanks to its high flexibility, MEPSA can easily be adapted and tailored
to different time series that are routinely obtained in many fields other than GRBs,
without having to modify the code itself.
\ref{sec:appendix} describes the output information provided by the code,
which is made publicly available
\footnote{\url{http://www.fe.infn.it/u/guidorzi/new_guidorzi_files/code.html}}.

\section{Algorithm's description}
\label{sec:desc}
MEPSA searches the input light curves for peaks by applying simultaneously a
multi--pattern set of excesses, i.e. a set of $N=39$ patterns.
The input time series must have uniform
sampling and must be affected by statistical uncorrelated Gaussian noise.
The series must be either background--subtracted, or removed of possible
trends, so that changes in the expected value should only be due to signal and
not to background.
For each bin of the light curve, let $r_i$ the rates in the $i$-th bin.
A given pattern $P_k$ ($k=1,\ldots,N$) consists of a fixed number of adjacent
bins around the given $i$-th bin: around $i$ there are a given $n_{k,l}$ leftward
bins (which temporally precede the $i$-th bin) and $n_{k,r}$ rightward bins (which
temporally follow the $i$-th bin). The pattern assigns each of its bins, except for
the $i$-th bin, a threshold $v_{k,j}$ ($j=1,\ldots,n_{k,l}+n_{k,r}$) in terms of
number of $\sigma$'s (where $\sigma$ is the statistical noise corresponding to
that bin). Pattern $P_k$ is then said to be fulfilled by bin $i$ when the following
conditions are simultaneously fulfilled:
\begin{equation}
\left\{
 \begin{array}{lr}
   r_i - r_j  \ge  v_{k,(j-i+n_{k,l}+1)}\,\sigma'_{ij} &  (j=i-n_{k,l},\ldots,i-1)\\
   r_i - r_j  \ge  v_{k,(j-i+n_{k,l})}\,\sigma'_{ij}   & (j=i+1,\ldots,i+n_{k,r})
   \end{array} \right.
   \label{eq:alg}
\end{equation}
where $\sigma'_{ij} = (\sigma^2_i + \sigma^2_j)^{1/2}$.
Each pattern has different numbers of leftward and rightward bins as well as different
threshold values for each of them.
For each bin $i$ the search is performed by applying simultaneously a set of 39 different
patterns ($k=1,\ldots,39$) and the $i$-bin is promoted to peak candidate if at least
one pattern is fulfilled. The complete set of threshold values $v_{k,j}$ currently
adopted is reported in Table~\ref{tab:pattern}.
Operatively, threshold values are not hard--coded, but are stored within an external
file: this allows users to disable existing patterns/enable new ones in a very
flexible way.

When the entire light curve has been screened, the whole procedure is repeated to
rebinned versions of the same curve, each time increasing the rebinning factor by one
up to a maximum value $F_{\rm reb,m}$ established by the user.
Moreover, for a rebinning factor of $F_{\rm reb}$ (integer) of the original light
curve, there are $F_{\rm reb}$ possible offsets: in fact, one can choose $F_{\rm reb}$
different starting bins and end up with as many different rebinned profiles.
For for a given $F_{\rm reb}$ rebinning factor, all the corresponding rebinned curves
are searched through the same algorithm.
The goal behind this is to identify peaks with very different SNR and/or
very different timescales.
The computational time scales as $F_{\rm reb,m}^3$, so unreasonably high values should be avoided.

Clearly, most of the peaks are expected to be detected in multiple searches. So a final
crosscheck is performed to make sure the same peak candidate is not going to be classified
repeatedly as a number of distinct peak candidates. This is done by comparing the peak times
together with the timescales associated with each peak time.
Finally, when the same peak candidate is detected at different timescales, the algorithm
selects the one with the (statistically most significant) highest value. This timescale is therefore defined
as ``detection timescale'' and hereafter is denoted with $\Delta t_{\rm det}$.
This automatically identifies the timescale at which the peak is detected at best; as such,
this can be taken as a reasonable proxy for the peak timescale itself.

The set of values chosen for the pattern thresholds had been determined after analysing
hundreds of GRB profiles detected with {\em CGRO}/BATSE \citep{Paciesas99}, {\em BeppoSAX}/GRBM
\citep{Frontera09}, and {\em Swift}/BAT \citep{Sakamoto11}.
As such, they were tailored to the GRB features themselves so as to ensure
an acceptable trade-off between the rate of true negative and that of false positive
detections.
In this respect, the algorithm in its current version (October 2014) has proved to be conservative,
as is shown in Sect.~\ref{sec:valid}.

\begin{table*}
\caption{Matrix of excess thresholds. Each line refers to a single pattern $P_k$,
identified by $k$. $n_{k,l}$ and $n_{k,r}$ are the numbers of leftward and rightward
bins, respectively, as in Eq.~(\ref{eq:alg}). Threshold values $v_{k,j}$
$(j=1,\ldots,n_{k,l}+n_{k,r})$ are given in the numbered columns.}
\label{tab:pattern}
\begin{tabular}{rccrrrrrrrrrr}
\hline\noalign{\smallskip}
$k$ & $n_{k,l}$ & $n_{k,r}$ & $v_{k,1}$ & $v_{k,2}$ & $v_{k,3}$ & $v_{k,4}$ & $v_{k,5}$ & $v_{k,6}$ & $v_{k,7}$ & $v_{k,8}$ & $v_{k,9}$ & $v_{k,10}$  \\
\hline\noalign{\smallskip}
  1 & 1 & 1  &   5.0 &   5.0 &    - & -   & -   & -   & -   & -   & -   & -\\ 
  2 & 1 & 2  &   5.0 &   1.0 &  5.0 & -   & -   & -   & -   & -   & -   & -\\
  3 & 1 & 3  &   5.0 &   4.8 &  2.0 & 3.5 & -   & -   & -   & -   & -   & -\\
  4 & 1 & 3  &   5.0 &   2.0 &  2.2 & 5.0 & -   & -   & -   & -   & -   & -\\
  5 & 2 & 1  &   5.0 &   1.0 &  5.0 & -   & -   & -   & -   & -   & -   & -\\
  6 & 2 & 2  &   5.0 &   2.0 &  2.0 & 5.0 & -   & -   & -   & -   & -   & -\\
  7 & 2 & 3  &   4.5 &   3.0 &  2.0 & 3.5 & 5.0 & -   & -   & -   & -   & -\\
  8 & 2 & 3  &   5.0 &   3.0 &  4.5 & 3.0 & 5.0 & -   & -   & -   & -   & -\\
  9 & 3 & 1  &   5.0 &   2.0 &  2.2 & 5.0 & -   & -   & -   & -   & -   & -\\
 10 & 3 & 1  &   5.0 &   4.5 &  1.5 & 5.0 & -   & -   & -   & -   & -   & -\\
 11 & 3 & 2  &   5.0 &   3.0 &  4.5 & 3.0 & 5.0 & -   & -   & -   & -   & -\\
 12 & 3 & 3  &   3.0 &   2.8 &  1.7 & 2.0 & 4.0 & 5.0 & -   & -   & -   & -\\
 13 & 3 & 3  &   5.0 &   4.5 &  2.0 & 2.0 & 4.0 & 2.5 & -   & -   & -   & -\\
 14 & 3 & 4  &   5.0 &   4.5 & -2.0 & 0.2 & 2.0 & 2.2 & 2.0 & -   & -   & -\\
 15 & 4 & 1  &   2.0 &   3.3 &  2.6 & 2.4 & 5.0 & -   & -   & -   & -   & -\\
 16 & 4 & 2  &   5.0 &   1.7 &  2.2 & 2.5 & 2.4 & 5.0 & -   & -   & -   & -\\
 17 & 4 & 3  &   5.0 &   4.6 &  0.8 & 0.4 & 1.4 & 1.3 & 5.0 & -   & -   & -\\
 18 & 4 & 3  &   5.0 &   3.5 &  2.0 & 3.0 & 3.0 & 3.5 & 4.5 & -   & -   & -\\
 19 & 4 & 3  &   3.4 &   1.8 &  1.2 & 3.4 & 1.2 & 3.8 & 5.0 & -   & -   & -\\
 20 & 4 & 3  &   5.0 &   2.1 &  2.2 & 3.0 & 1.8 & 3.4 & 5.0 & -   & -   & -\\
 21 & 4 & 5  &   4.9 &   3.5 &  2.0 & 1.8 & 1.8 & 2.9 & 3.3 & 3.2 & 3.2 & -\\
 22 & 5 & 2  &   3.4 &   2.8 &  2.0 & 3.4 & 3.5 & 3.4 & 5.0 & -   & -   & -\\
 23 & 5 & 3  &   3.0 &   2.8 &  3.5 & 0.2 & 1.0 & 1.9 & 4.3 & 5.0 & -   & -\\
 24 & 5 & 3  &   1.7 &   2.4 &  1.4 & 2.0 & 1.0 & 2.2 & 4.0 & 5.0 & -   & -\\
 25 & 5 & 4  &   3.4 &   3.8 &  4.0 & 3.0 & 1.5 & 0.3 & 1.2 & 2.7 & 4.0 & -\\
 26 & 5 & 4  &   2.2 &   3.9 &  2.2 & 3.4 & 0.7 & 3.1 & 2.2 & 1.6 & 1.7 & -\\
 27 & 5 & 4  &   1.5 &   2.6 &  2.4 & 2.5 & 1.0 & 1.5 & 2.5 & 2.5 & 4.8 & -\\
 28 & 5 & 4  &   4.5 &   1.4 &  4.0 & 1.9 & 1.1 & 1.9 & 2.8 & 3.8 & 3.0 & -\\
 29 & 5 & 5  &   0.5 &  -1.8 & -0.1 & 2.7 & 3.8 & 4.0 & 2.5 & 1.5 & 3.8 & 4.0\\
 30 & 5 & 5  &   3.5 &   4.0 &  2.5 & 2.0 & 1.0 & 0.7 & 2.0 & 2.1 & 3.1 & 2.8\\
 31 & 5 & 5  &   5.0 &   5.0 &  5.0 & 4.0 & 2.3 & 0.5 & 1.4 & 3.0 & 3.0 & 2.7\\
 32 & 5 & 5  &   2.3 &   3.6 &  2.6 & 0.9 & 1.8 & 2.1 & 2.9 & 4.1 & 3.6 & 2.7\\
 33 & 5 & 5  &   3.0 &   4.0 &  2.5 & 2.8 & 0.7 & 2.4 & 3.3 & 4.0 & 4.5 & 3.0\\
 34 & 5 & 5  &   3.1 &   2.8 &  3.4 & 1.2 & 1.4 & 2.8 & 2.0 & 3.6 & 3.2 & 3.2\\
 35 & 5 & 5  &   3.4 &   3.6 &  3.0 & 1.6 & 0.6 & 2.3 & 2.0 & 0.8 & 3.2 & 3.2\\
 36 & 2 & 2  &   3.0 &   4.0 &  4.0 & 3.0 & -   & -   & -   & -   & -   & -\\
 37 & 3 & 3  &   2.5 &   3.5 &  3.5 & 3.5 & 3.5 & 2.5 & -   & -   & -   & -\\
 38 & 3 & 3  &   3.0 &   4.0 &  0.0 & 0.0 & 4.0 & 3.0 & -   & -   & -   & -\\
 39 & 4 & 4  &   3.0 &   4.0 &  0.0 & 0.0 & 0.0 & 0.0 & 4.0 & 3.0 & -   & -\\
\noalign{\smallskip}\hline\noalign{\smallskip}
\noalign{\smallskip}\hline
\end{tabular}
\end{table*}
%

\subsection{Li--Fenimore algorithm}
\label{sec:LF}
For comparison purposes, we considered the peak search algorithm proposed
by \citet{Li96} (hereafter LFA), that has become popular in the GRB literature
\citep{NakarPiran02a,DragoPagliara07,Bhat12} as well as in other fields
\citep{Feng99,Liu04}.

\paragraph{(classical) LFA}
According to LFA prescription, a local maximum at $i$--th bin is promoted to
peak candidate when two nearby valleys at $j$--th and $k$--th bins are found,
so that $(r_i-r_{j,k})\ge n\,\sigma_i$ ($n=5$), with no other higher peak lying
between the same two valleys. Hereafter, LFA in its original formulation
will be referred to as ``classical LFA'' or simply LFA to distinguish it
from a slightly modified version of the same algorithm we conceived.

\paragraph{Conservative LFA (cLFA)}
The so--called conservative LFA (hereafter, cLFA) works in the same way as LFA,
except for the condition is slightly but importantly different:
$(r_i-r_{j,k})\ge n\,(\sigma^2_i+\sigma^2_{j,k})^{1/2}$ ($n=5$).
In principle, this accounts for the variance of excesses in a statistically
more correct way. For many realistic cases of astrophysical interest, time profiles
are nearly homoscedastic (i.e., all $\sigma$'s are comparable with each other),
so cLFA essentially imposes thresholds as high as $\sim\sqrt{2}$ times those of LFA.
Its expected robustness (greater than that of LFA against the misclassification of
statistical fluctuations as false positives), explains the ``conservative'' labelling.

\section{Tests and validation}
\label{sec:valid}
We tested MEPSA by means of simulated time profiles that were used to evaluate
the following characterising features:
\begin{enumerate}
\item false positive (FP) rate;
\item true positive (TP) or, equivalently, true negative (TN) rate.
\end{enumerate}
Following standard naming conventions, FPs denote statistical fluctuations
which do not correspond to any real peak and are misclassified by MEPSA as
genuine peak candidates. The higher the FP rate, the less pure the
sample of peak candidates.
TPs are instead real peaks which are correctly identified as such; they
complement the number of TNs, which are instead real peaks being missed.
The higher the TP rate, the more complete the peak candidate sample.

The ideal algorithm has a null FP rate and 100\% TP rate. In practice,
the two criteria compete with each other, so that only a compromise is feasible.
The best trade--off is to be tailored to the goal of a given experiment,
depending on which, between purity and completeness of the sample, is more crucial.
Concerning the problem of peak identification, unless one has specific
requirements, in most cases purity is more important. On the other hand
the capability to identify dim peaks can be worth pursuing e.g., when
one aims at studying waiting time distributions, provided that it costs
a relatively low number of FPs.

\subsection{False Positive rate}
\label{sec:FP}
We simulated a number of constant time profiles affected by Gaussian uncorrelated
noise and applied the three algorithms. We assumed a fixed bin time of 64~ms,
which is the typical resolution of GRB experiments such as BATSE, and is
significantly shorter than $0.6$--$1$~s, that is the characteristic variability
timescale range observed in GRB profiles \citep{Margutti11c}.
We generated two different groups of curves:
\begin{description}
\item[group 1] $N=300$ time profiles with $N_b=5000$ bins each. We generated Poisson
distributed counts with an expected rate $r_e=1000$~counts/bin. Such high-counting
regime ensures that rates are normally distributed according to a $N(r_e,\sqrt{r_e})$.
These values are representative of background counts for scintillators such as
those which flew aboard {\em CGRO} or {\em BeppoSAX} operating from a few dozens
up to several hundreds keV.
\item[group 2] $N=100$ time profiles with $N_b=15000$ bins each. We generated Gaussian
distributed rates according to $N(0,\sigma_i)$, where $\sigma_i$ were taken from
a typical mask-weighted light curve of {\em Swift}/BAT in the 15--150~keV band.\footnote{We
took the time profile of GRB\,100814A \citep{Krimm10}, which consists
of a complex superposition of a number of pulses with different durations.}
\end{description}
Each group totals to $1.5\times10^6$~bins purely affected by statistical noise
with no real structure. Table~\ref{tab:FP} reports the number of FPs for each
group and for each algorithm.
\begin{table}
\caption{Number of FPs detected by each algorithm out of two groups of simulated curves without peaks.
The corresponding fractions (out of $1.5\times10^{6}$ scanned bins) are among brackets.}
\label{tab:FP}
\begin{tabular}{lccc}
\hline\noalign{\smallskip}
G & MEPSA &  cLFA &  LFA\\
\hline\noalign{\smallskip}
1     &    $20\, (1.3\times10^{-5})$ &  $107\, (7.1\times10^{-5})$ & $5263\, (3.5\times10^{-3})$\\
2     &    $36\, (2.4\times10^{-5})$ &  $162\, (1.1\times10^{-4})$ & $7085\, (4.7\times10^{-3})$\\
\hline\noalign{\smallskip}
\end{tabular}
\end{table}
\begin{figure}[!h]
  \begin{center}
  \includegraphics[width=0.45\textwidth]{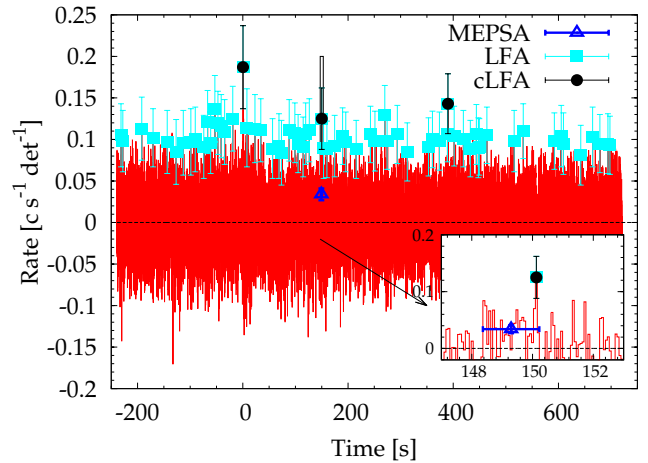}
  \caption{Example of simulated featureless profile affected by
    uncorrelated Gaussian noise. MEPSA, LFA, cLFA found 1, 65, 3 FPs
    peaks, respectively. A close-in of the only MEPSA FP is shown
    in the inset. This time profile mimics the background of a typical
    mask-weighted curve by {\em Swift}/BAT.}
  \end{center}
  \label{fig:example_FP}
\end{figure}
As an illustrative example Figure~\ref{fig:example_FP} shows a group--2
light curve together with the FPs identified by each algorithm.
MEPSA has the lowest FP rate, $\sim1$--$2\times10^{-5}$ FP/bin, which
is equivalent to a $4.2$--$4.4\,\sigma$ (Gaussian) significance threshold.
cLFA has a significantly worse FP rate, about 4--5 times
as high, corresponding to $3.9$--$4.0\,\sigma$ (Gaussian), whereas
LFA is by far the worst, with a rate of $3$--$5\times10^{-3}$~FP/bin,
corresponding to $2.8$--$2.9\,\sigma$ (Gaussian).
The ratio between LFA and cLFA significance Gaussian thresholds is
indeed compatible with $\sqrt{2}$, as expected (Sect.~\ref{sec:LF}).

\begin{figure}
  \includegraphics[width=0.45\textwidth]{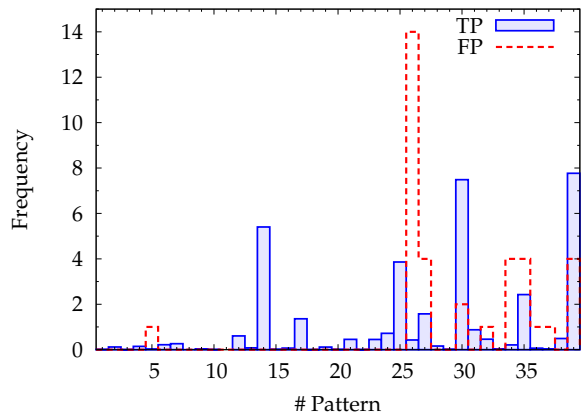}
  \caption{MEPSA FP (dashed) and TP (solid) spectra, i.e.,
number of events as a function of the triggering pattern.
The TP spectrum has been downsized so as to have the same area
as the FP one.}
  \label{fig:crit}
\end{figure}
In the case of MEPSA, we also studied the FP rate as a function of the
triggering pattern through what can be considered as a ``FP spectrum''.
The dashed histogram in Figure~\ref{fig:crit} shows how the 36 FPs
detected in group 2 profiles distribute among the 39 patterns.
Clearly, pattern 26 has the highest FP rate and thus
appears to be the weakest pattern in this respect.
The FP spectrum is compared with the TP one obtained from the simulated
peaks described in Sect.~\ref{sec:TP}.

\begin{figure*}[!ht]
  \includegraphics[width=1.0\textwidth]{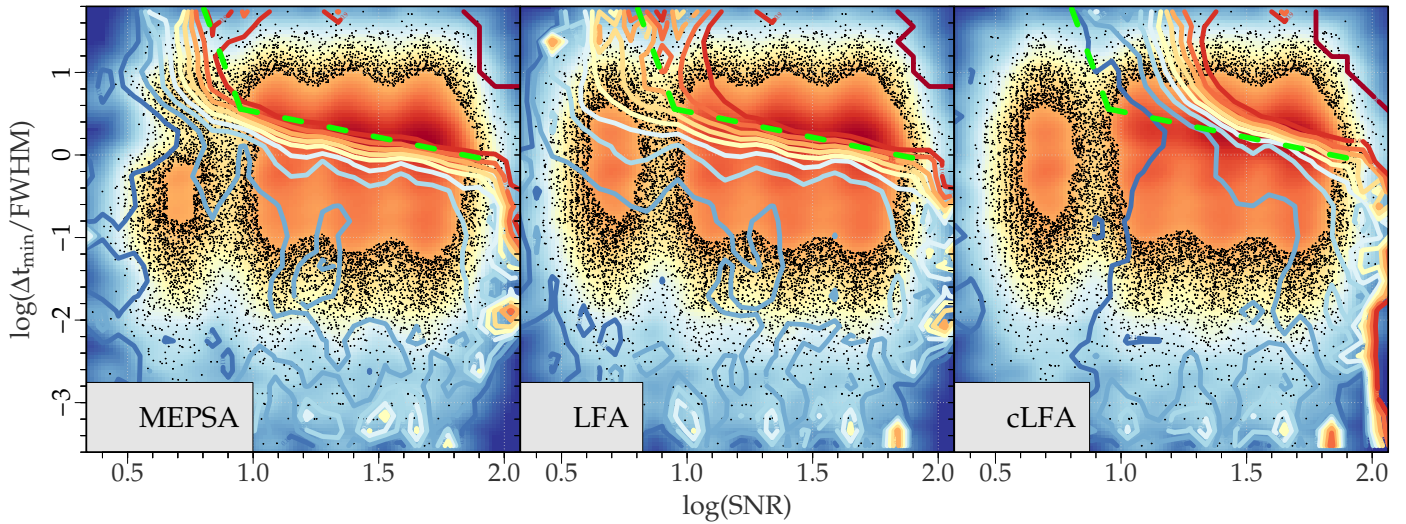}
  \caption{Peak detection efficiency in the SNR--separability plane for MEPSA ({\em left}), 
    LFA ({\em mid}), and cLFA ({\em right}). Different Contour levels (from cold to hot colours)
    correspond to ten different, equally spaced efficiency levels from 0\% to 100\% with increasing order.
    The MEPSA 90\% contour level is approximately described with two connected power-laws (dashed line) and
    is shown in all panels for comparison. Dots and background colours represent the bivariate
    distribution of the {\em detected} peaks.}
  \label{fig:eff_MEPSA}
\end{figure*}
%

\subsection{True Positive rate}
\label{sec:TP}
Starting from the same template of {\em Swift}/BAT mask--weighted background
time profile used to build group 2, we simulated 150 curves populated with
fast--rise exponential decay (FRED) pulses assuming the model by \citet{Norris96}.
We assumed a fixed value for the peakedness given by the average value found
in real GRBs, $\nu=1.5$ \citep{Norris96}.
This corresponds to a pulse shape which lies between a pure exponential and a Gaussian profile.
We also assumed fixed rise and decays times, $\sigma_{\rm r}=1$ and $\sigma_{\rm d}=3$~s,
respectively, in agreement with what is typically observed in GRB time profiles \citep{Norris96}.
In this mathematical formulation, the corresponding full width at half maximum (hereafter,
FWHM) amounts to $(\ln{2})^{1/\nu}\,(\sigma_{\rm r} + \sigma_{\rm d})$.

As will be shown in the following, the relevant parameter for the peak detection of a given
pulse is the ratio between its lowest adjacent waiting time $\Delta t_{\rm min}$ and its FWHM,
rather than the absolute value of the FWHM itself. 
For a given $i$--th pulse peaking at $t_{{\rm p},i}$, the lowest adjacent waiting time is
defined as $\Delta t_{{\rm min},i}={\rm min}(t_{{\rm p},i}-t_{{\rm p},i-1}, t_{{\rm p},i+1}-t_{{\rm p},i})$.
This explains our choice of adopting a fixed FWHM for the simulated pulses,
since we varied the waiting time distribution. The higher the ratio, the more easily
the pulse can be recognised as a separate entity from its surrounding siblings.
We therefore define the {\em separability} $s_i$ of a given pulse $i$, as 
\begin{equation}
s_i = \frac{\Delta t_{{\rm min},i}}{{\rm FWHM}_i}\;.
\label{eq:sep}
\end{equation}
The more two adjacent pulses overlap, the lower their individual separabilities, and
correspondingly harder for a given algorithm is to identify them as two separate entities.
The peakedness, which determines the shape of the pulse profile, is expected
to have a minor weight as far as the peak detection is concerned, provided that extreme
and nonphysical values are not considered. This justifies our choice of assuming a typical,
fixed value for it, and our choice of exploring more in detail the effects of other
more crucial parameters, such as the SNR of the pulse, i.e. the ratio between the total
area (or fluence) and its statistical uncertainty, and the ratio between the time intervals
mentioned above.

Different pulses within the same simulated curve were generated assuming an exponential
distribution for the waiting times, i.e. the case of a memoryless process with a constant
expected Poisson rate of pulses per unit time. We assumed a range for the expected rates
of pulses from $1/40$ up to $1/20$ pulses~s$^{-1}$. Peak intensities were assumed so as
to cover the SNR range from $2$ all the way up to $100$.
The nature of the waiting time distribution (an exponential in this case) is not relevant
to our goal; rather, the aim is covering as much as possible the SNR--separability plane
and monitoring the algorithms' efficiency as a function of both variates.

Overall, $89\,540$ pulses were generated spanning the ranges $0.5\lesssim\log{({\rm SNR})}\lesssim 2.0$
and $-3\lesssim\log{s}\lesssim 2$.

\subsubsection{Efficiency}
\label{sec:eff}
We split the SNR--$s$ plane in 30 different boxes and for each of them we calculated
the fraction of identified peaks over the total number of pulses. Left--hand panel in
Figure~\ref{fig:eff_MEPSA} shows the MEPSA efficiency. 
The 90\% contour level can approximately be described with a double piece-wise power--law
(dashed line in Fig.~\ref{fig:eff_MEPSA}), whose equation is
\begin{equation}
\log{s_{0.9}({\rm SNR})} = \left\{
 \begin{array}{lr}
   -8.28\,\log{({\rm SNR})} + 8.42 &  (\log{({\rm SNR})<0.95})\\
   -0.63\,\log{({\rm SNR})} + 1.15 & (\log{({\rm SNR})\ge 0.95})
   \end{array} \right.\;.
   \label{eq:eff_MEPSA}
\end{equation}
\begin{figure*}[!ht]
  \includegraphics[width=0.45\textwidth]{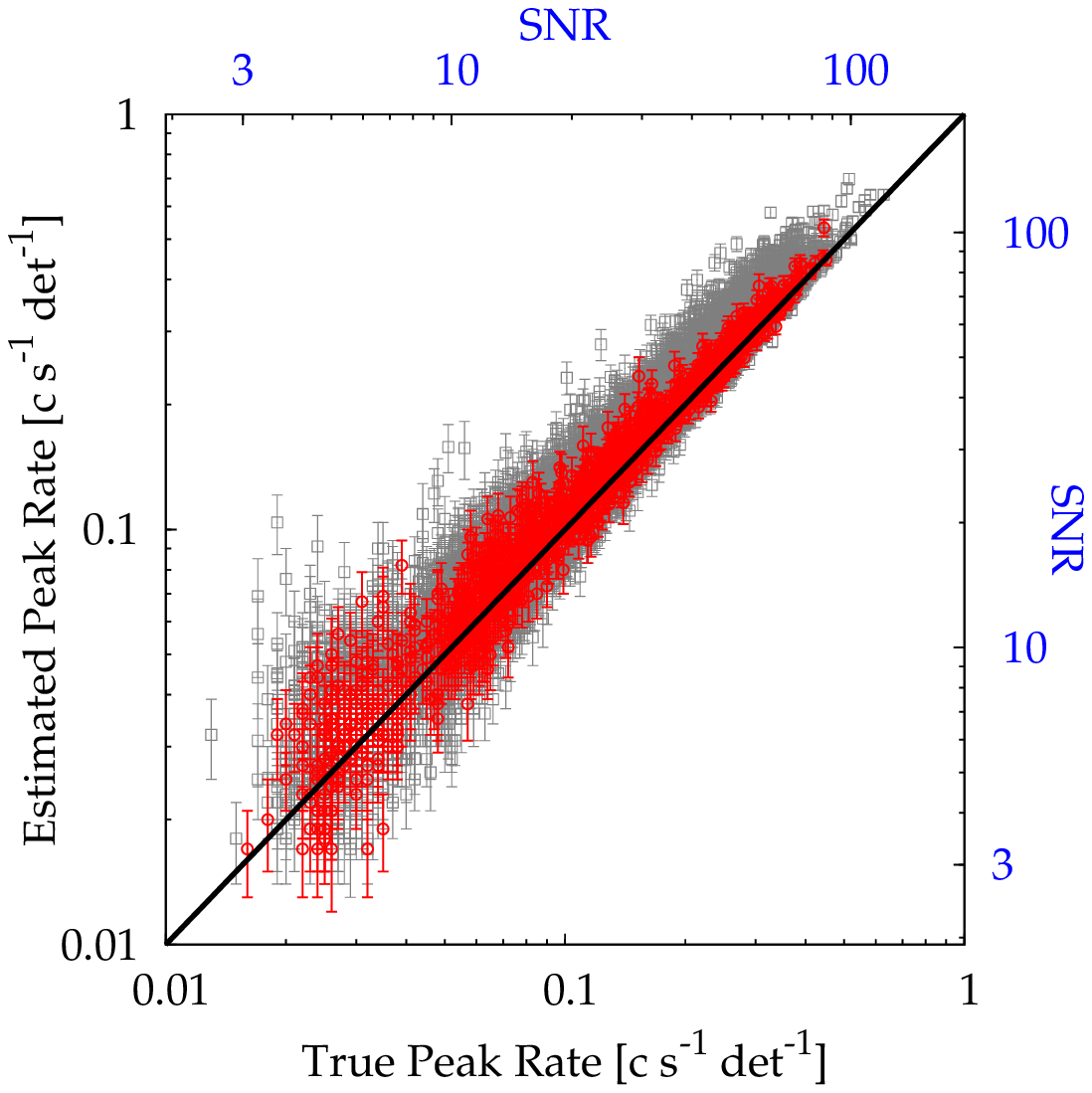}
  \includegraphics[width=0.55\textwidth]{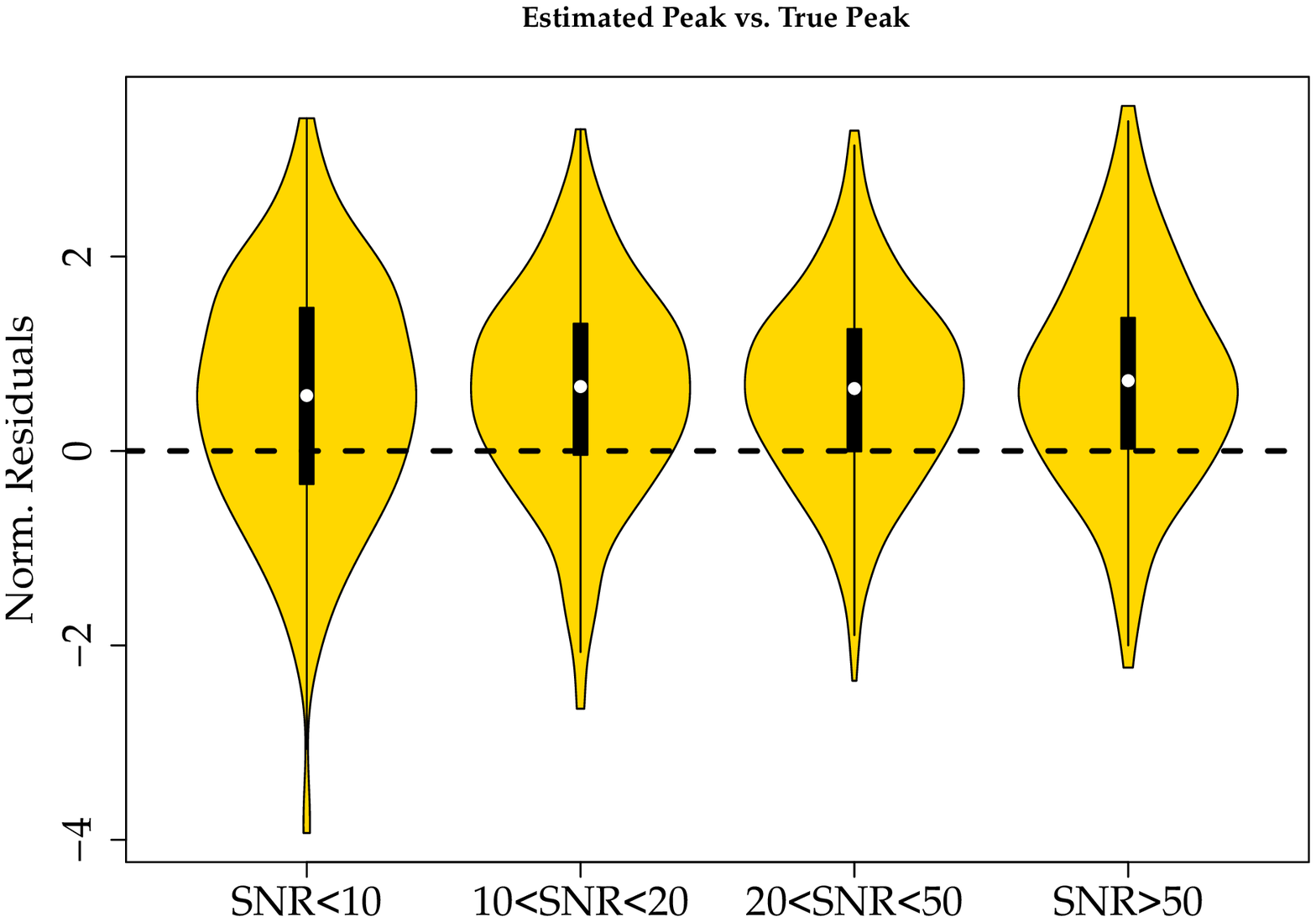}
  \caption{{\em Left panel}: MEPSA estimated vs. true peak rates for weakly separated
    ($1<s<10$, grey) and well separated pulses ($s>10$, red). The solid line shows equality.
    {\em Right panel}: corresponding violin plot of normalised residuals of estimated vs.
    true values for different classes of SNR of well separated pulses.}
  \label{fig:A_vs_R}
\end{figure*}
Whenever the condition $s({\rm SNR})\ge s_{0.9}({\rm SNR})$ is fulfilled, MEPSA efficiency is 90\% at least.
Worth noting are the following properties:
\begin{itemize}
\item efficiency is very high in the top right region of well separated,
  high contrast pulses, as one expects;
\item when $\log{s}<-0.4$, i.e. $s<0.4$, pulses can hardly ($<$10--20\%)
be identified as separate pulses, regardless of the SNR;
\item at a given separability, efficiency slightly improves for higher SNR;
\item when SNR drops below 4--5, efficiency drops as well almost regardless of separability;
\item at fixed SNR$>$4--5, efficiency drops from 90\% to 50\% when separability decreases
from $s_{0.9}$(SNR) by a factor of $\sim10^{-0.2}\approx0.6$.
\end{itemize}

Likewise, we studied of efficiency in the SNR--$s$ plane for the other two algorithms.
Mid and right--hand panels in Figure~\ref{fig:eff_MEPSA} show the results for
LFA and cLFA, respectively: the dashed line in both plots is the same as in right--hand
panel and described by Eq.~(\ref{eq:eff_MEPSA}) for comparison.
At given separability values, LFA has comparable efficiency values as long as
SNR$\gtrsim15$. However, at SNR~$<15$ LFA efficiency becomes remarkably worse than MEPSA
(e.g., 60\% vs. 90\%, at SNR~$\sim8$ and $s>3$). This proves that MEPSA has a higher TP
(lower TN) rate than LFA in low-intermediate SNR range, i.e. 4--5~$<$~SNR~$<$~15.

Compared with MEPSA and LFA, the TP rate of cLFA is the lowest (highest TN rate),
since the $> 90$\% efficiency region shrinks along both SNR and $s$, as shown in the right--hand
panel of Fig.~\ref{fig:eff_MEPSA}. This is no wonder, because it is the downside
of a relatively low FP rate, typical of a more selective algorithm.

\subsubsection{Accuracy of peak rate estimates}
\label{sec:rate}
We investigated the accuracy of different algorithms in estimating peak intensities
or rates, something that can be done only for the TP peaks.
Left--hand panel of Figure~\ref{fig:A_vs_R} shows MEPSA--estimated vs. true peak rates for two
different classes of separability $s$: the mildly separated pulses ($1<s<10$, grey), and
the well separated ones ($s>10$, red).
The mildly separated peak rates tend to scatter more significantly above equality
than well separated ones do: the reason behind this is that more overlapping pulses
are more likely to be identified as a single peak, whose estimated rate is therefore
the sum of different pulses' contributions. While the scatter around equality seems
to enhance in the low--SNR (or low--rate) end of the distribution, the corresponding
uncertainties increase as well.
To better understand whether the accuracy of a MEPSA--estimated peak
rate depends on SNR or, equivalently, on the value of the rate itself,
we show in the right--hand panel of Fig.~\ref{fig:A_vs_R}
the violin plot of the normalised scatter, i.e. the difference
between true and estimated values normalised by the estimated uncertainty, as a function
of peak rate. We considered the sample of well separated pulses ($s>10$) only.
There is no significant trend with peak rate: the variance of distribution
remains essentially unchanged throughout the spanned range, and so does the mean value.
In all cases the null value is within $1\,\sigma$ of the distribution.
All mean values are above zero, suggestive of a small bias that tends to overestimates
the peak rate, even though it is still within uncertainties. A possible explanation for
that is that MEPSA searches the peak through all the possible binnings and shifts,
with the result of a slight bias towards positive statistical fluctuations.

\begin{figure}
  \includegraphics[width=0.45\textwidth]{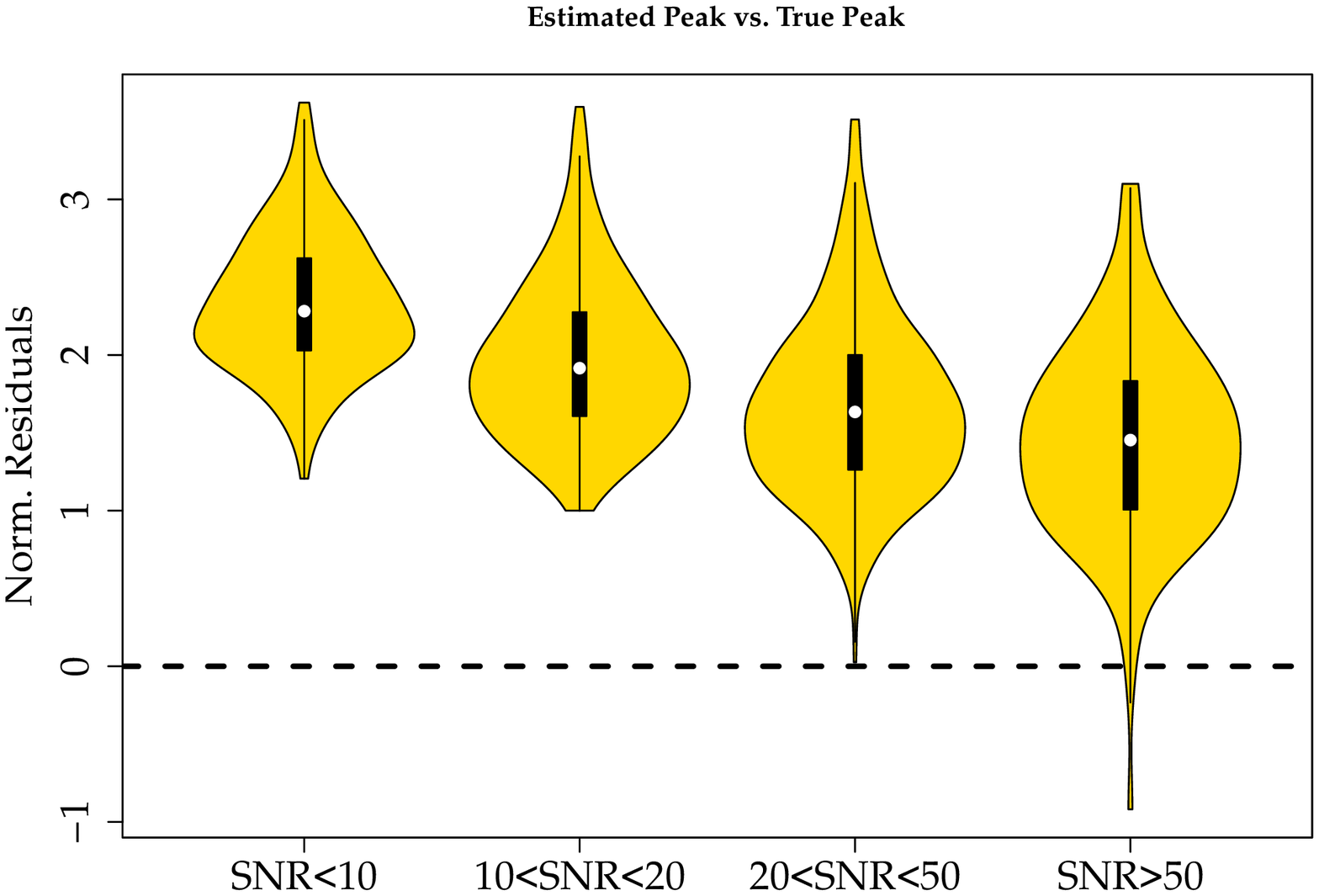}
  \includegraphics[width=0.45\textwidth]{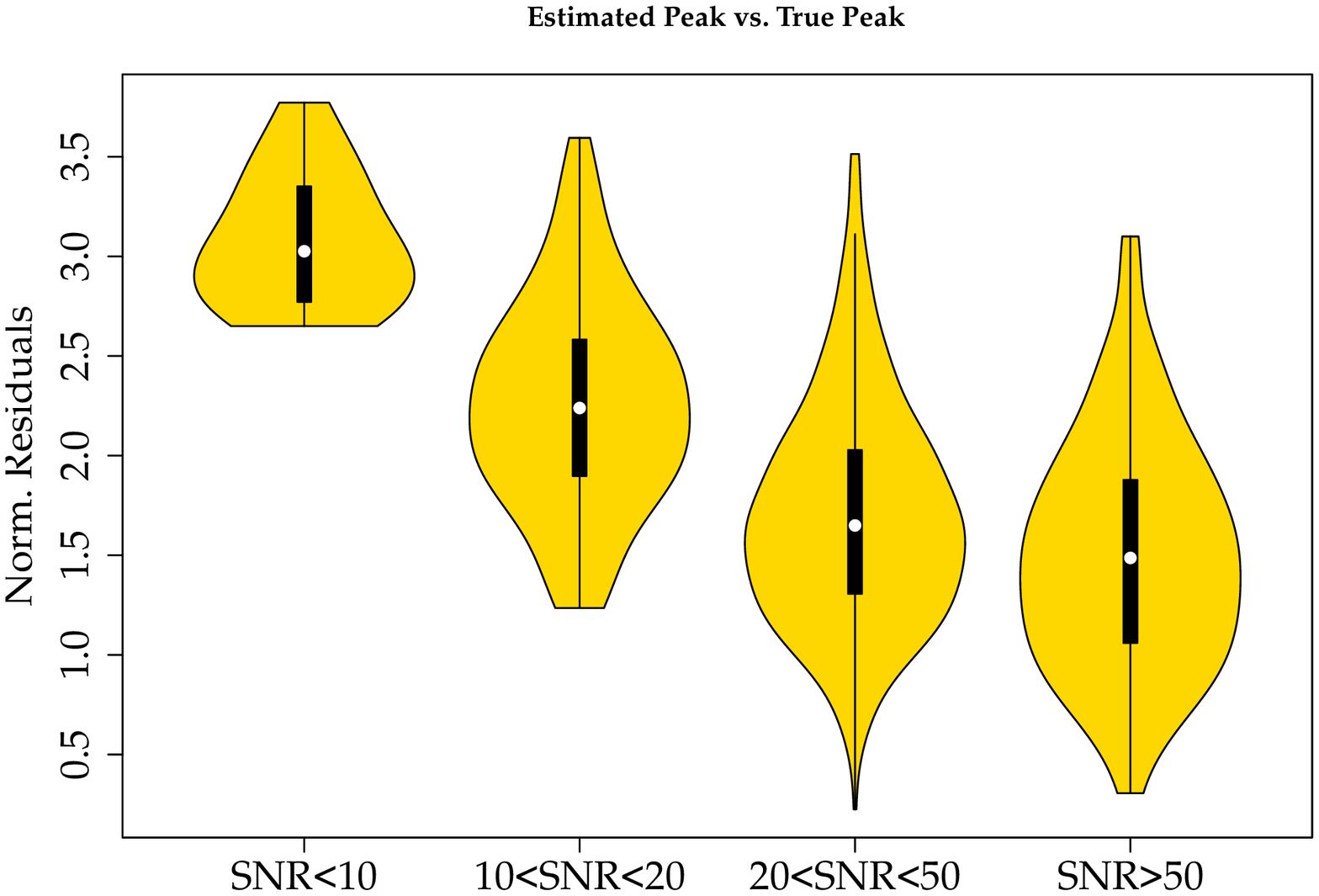}
  \caption{Violin plots of the normalised residuals for both versions
    of LF algorithms: LFA ({\em top}) and cLFA ({\em bottom}),
    analogously to Fig.~\ref{fig:A_vs_R}.
    Both LFA and cLFA provide significantly more biased estimates
    of the true peak rates than our algorithm does.}
  \label{fig:A_vs_R_LF}
\end{figure}

Figure~\ref{fig:A_vs_R_LF} shows the analogous violin plots for the normalised residuals
for LFA and cFLA, respectively, for the identified well separated pulses.
In both cases, the accuracy is remarkably worse at low SNR values, where peak rate
estimates become crucially biased by statistical
fluctuations that overestimate as much as up to $\sim3\,\sigma$.
It is worth noting that in this respect cLFA is even worse on average.
%

\subsubsection{MEPSA proxy for the peak FWHM}
\label{sec:DeltaTbin}
Both MEPSA and LFA algorithms do not assume any specific shape for the peaks.
As non--parametric methods, they have the benefit of being applicable to a broad
variety of time series characterised by peaks. The downside is that peak FWHM
must be estimated from the data themselves and not through fitting
parameters that are linked to specific peak models.
LFA algorithms provide no direct information about peak width, the only bare
proxy being the times of the two adjacent valleys surrounding a given peak.

\begin{figure}
  \includegraphics[width=0.45\textwidth]{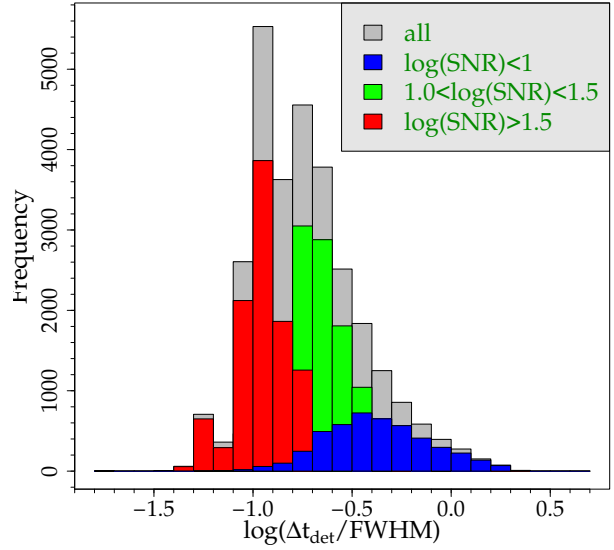}
  \caption{Distribution of the ratio between MEPSA detection timescale
    $\Delta t_{\rm det}$, used as a proxy for the pulse FWHM, and the FWHM itself.
    This is done separately for different SNR classes.}
  \label{fig:DeltaTbin}
\end{figure}

Analogously, a detection timescale $\Delta t_{\rm det}$ is associated to each
MEPSA peak (Sect.~\ref{sec:desc}).
As in the case of LFA valleys, the detection timescale
is affected by the peak SNR: at low SNR values it tends to bin up the light
curve as much as possible to reach the required statistical excess to trigger
MEPSA. On the other side, at high SNR the peak already triggers some MEPSA
patterns at low timescales and at longer timescale, although the SNR increases,
the average peak rate estimate likely decreases. All this turns into favouring
the short timescales at high SNR and the other way around at low SNR.
This is indeed what is observed when one studies the distribution of the
ratio between detection timescale and FWHM for three different SNR ranges,
as shown by Fig.~\ref{fig:DeltaTbin}.
Looking at the mean values and scatters of each distribution, we conclude
that, on average,
\begin{itemize}
\item at $\log{({\rm SNR})}>1.5$,  it is FWHM$\approx10\,\Delta_{\rm det}$,
\item at $1.0<\log{({\rm SNR})}<1.5$, it is FWHM$\approx5-6\,\Delta_{\rm det}$,
\item at $\log{({\rm SNR})}<1.0$,  it is FWHM$\approx2.5\,\Delta_{\rm det}$,
\end{itemize}
with a factor of $\approx$2--3 uncertainty.

\subsubsection{MEPSA proxy for the peak SNR}
\label{sec:snr}
Similarly to peak FWHM discussed in Sect.~\ref{sec:DeltaTbin}, MEPSA does
not provide a direct estimate of the peak SNR, to calculate which one
should integrate over the entire pulse profile or, equivalently, over
its FWHM and correct it by a given factor.
As it was shown in Sect.~\ref{sec:DeltaTbin}, MEPSA can only roughly
estimate the FWHM, and so does it for the SNR, too.
A SNR proxy is yielded by the estimated SNR, SNR$_{\rm est}$, which
is calculated by MEPSA over the detection time $\Delta t_{\rm det}$.

\begin{figure}
  \includegraphics[width=0.45\textwidth]{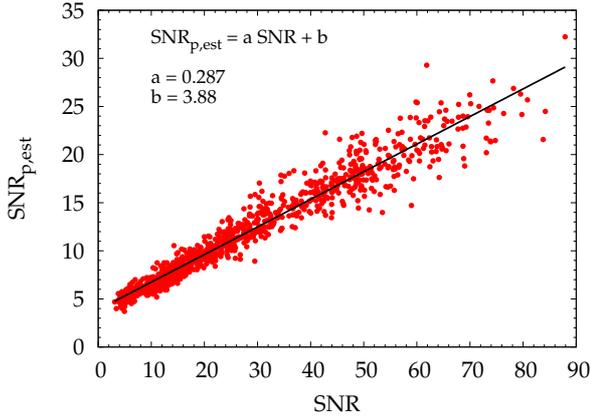}
  \caption{Relation between estimated and true SNR for a sample of
    well separated ($s>10$) peaks that have been identified by MEPSA.}
  \label{fig:snr}
\end{figure}

We studied the relation between the estimated and true SNR's for the
sample of well separated simulated peaks and obtained that, regardless
of the scatter, the relation can be approximately linearly described as,
\begin{equation}
  {\rm SNR}_{\rm est}\ \approx\ 0.29\,{\rm SNR} + 3.9\;,
  \label{eq:snr}
\end{equation}
as displayed by Figure~\ref{fig:snr}.

Practically, starting from the values obtained by MEPSA for the
SNR$_{\rm est}$ and $\Delta t_{\rm det}$ of a given peak, one could
use Eq.~(\ref{eq:snr}) to estimate the true SNR, and use it
to roughly estimate its FWHM using the approximate relations
of Sect.~\ref{sec:DeltaTbin}. This, in turn, allows one to place
the peak in the separability--FWHM plane shown in left--hand
panel of Fig.~\ref{fig:eff_MEPSA} and estimate the efficiency--corrected
rate of such peaks in the time series.

\section{Discussion and Conclusions}
\label{sec:disc}
We presented MEPSA, multiple excess peak search algorithm, which searches
for peaks by applying a mask of multi--excess patterns to an evenly spaced,
background--subtracted (or detrended) time series, which is thought to be
affected by statistical uncorrelated Gaussian noise.
This is often the case also for photon counting detectors operating in
the high counting regime. We compared its performance against the popular
and widely adopted LFA as well as with a slightly more conservative version
of the same, under two complementary aspects: the false positive and the
true positive rates. In either case MEPSA is more reliable, showing a lower FP
rate ($\lesssim2\times10^{-5}$~bin$^{-1}$) as well as a higher true positive one,
especially at low SNR ($\sim$ 4--5).
We showed that MEPSA efficiency most crucially depends on the combination
of separability, defined as the ratio between the lowest temporal separation
from adjacent peaks and the FWHM of a given peak, and SNR.
At SNR$<$4--5, efficiency significantly drops. This is also the case when adjacent
peaks overlap non--negligibly, i.e. when separability drops below 1, with little
but significant dependence on SNR, that we modelled with a double power--law
in the separability--SNR plane. MEPSA also yields some proxies to characterise
FWHM and SNR of pulses and we described a quick way to do that.

Although the motivation originally sprang up from GRB time profiles,
its applicability extends to other similar fields of high--energy astrophysics as well as
solar X--ray flares, and, more in general, whenever the applicability requirements
on the input time series are fulfilled.
The search algorithm is decoupled from the mask of multiple excesses being searched.
This property makes it particularly flexible, so that users possibly interested
in events other than GRBs can quickly modify and optimise the mask of multiple 
excesses, disable existing patterns and/or enable new ones, once these have been
tested through simulations or independent data sets.
The highly--portable C code is made publicly available so as to encourage a
broad optimisation through other kinds of astrophysical time series of interest.

\section*{Acknowledgements}
The author acknowledges support by PRIN MIUR project on ``Gamma Ray Bursts: from
progenitors to physics of the prompt emission process'', P.~I. F. Frontera (Prot. 2009 ERC3HT).
The author is also grateful to Adriano Baldeschi for useful discussions about the
MEPSA characterisation.

\appendix

\section{Information provided by MEPSA}\label{sec:appendix}
This section presents the information provided by MEPSA for each peak candidate.
Table~\ref{tab:output} shows an example.
Field names and header line are the same as what is printed to standard output
by MEPSA.
\begin{table*}
\caption{Information provided by MEPSA. Each line refers to each peak candidate.}
\label{tab:output}
\begin{tabular}{crrrrrrrrr}
\hline\noalign{\smallskip}
Peak& RebF& BinPhase &    PeakT &    BinT &    PeakR &   EPeakR  &   SNR   &   Criterium& Nadiac\\
\hline\noalign{\smallskip}
  1 &  35 &  34  &    -12.408 &   2.240  &      0.04405 &  0.00612 &   7.20  &  25 & 9\\
  2 &  11 &   6  &      1.800 &   0.704  &      0.07064 &  0.01129 &   6.25  &  30 &10\\
\hline\noalign{\smallskip}
\end{tabular}
\end{table*}
\begin{enumerate}
\item {\em Peak}: ordinal number of the peak candidate;
\item {\em RebF}: rebinning factor chosen by MEPSA of the original time series for the peak candidate;
\item {\em BinPhase}: binning phase (from 0 to RebF$-1$) chosen by MEPSA;
\item {\em PeakT}: peak time;
\item {\em BinT}: detection timescale, $\Delta t_{\rm det}$, which is given by the original time resolution
  of the time series multiplied by RebF;
\item {\em PeakR}: peak rate estimate (same units as the original time series);
\item {\em EPeakR}: error on {\em PeakR};
\item {\em SNR}: SNR$_{\rm est}$;
\item {\em Criterium}: triggered pattern (Sect.~\ref{sec:desc});
\item {\em Nadiac}: number of adjacent bins involved in the triggered pattern identified by {\em Criterium}.
\end{enumerate}



\end{document}